# Control of quantum electrodynamical processes by shaping electron wavepackets

Liang Jie Wong[1,*], Nicholas Rivera[2], Chitraang Murdia[2], Thomas Christensen[2],

John D. Joannopoulos[2], Marin Soljačić[2] and Ido Kaminer[3,*]

[1]School of Electrical and Electronic Engineering, Nanyang Technological University, 50 Nanyang Ave, Singapore 639798, Singapore.
[2]Department of Physics, Massachusetts Institute of Technology, Cambridge, MA 02139, USA.
[3]Department of Electrical Engineering, Technion, Haifa 32000, Israel.

[*]Email: liangjie.wong@ntu.edu.sg, kaminer@technion.ac.il

Abstract: Fundamental quantum electrodynamical (QED) processes such as spontaneous emission and electron-photon scattering encompass a wealth of phenomena that form one of the cornerstones of modern science and technology. Conventionally, calculations in QED and in other field theories assume that incoming particles are single-momentum states. The possibility that coherent superposition states, i.e. "shaped wavepackets", will alter the result of fundamental scattering processes is thereby neglected, and is instead assumed to sum to an incoherent (statistical) distribution in the incoming momentum. Here, we show that free-electron wave-shaping can be used to engineer quantum interferences that alter the results of scattering processes in QED. Specifically, the interference of two or more pathways in a QED process (such as photon emission) enables precise control over the rate of that process. As an example, we apply our concept to Bremsstrahlung, a ubiquitous phenomenon that occurs, for instance, in X-ray sources for state-of-the-art medical imaging, security scanning, materials analysis, and astrophysics. We show that free electron wave-shaping can be used to tailor both the spatial and the spectral distribution of emitted photons, enhancing their directionality and monochromaticity, and adding more degrees of freedom that make emission processes like Bremsstrahlung more versatile. The ability to tailor the spatiotemporal attributes of photon emission via quantum interference provides a new degree of freedom in shaping radiation across the entire electromagnetic spectrum. More broadly, the ability to tailor general QED processes through the shaping of free electrons opens up new avenues of control in processes ranging from optical excitation processes (e.g., plasmon and phonon emission) in electron microscopy to free electron lasing in the quantum regime.

Free-electron-driven technologies lie at the heart of modern science and engineering, from X-ray tubes used in medical imaging, industrial quality inspection and security scanning, to electron microscopes that can capture fundamental phenomena with sub-angstrom [1,2] and sub-picosecond resolution [3,4]. The useful range of electron kinetic energies runs the gamut from non-relativistic energies, as low as 50 eV in applications like coherent low-energy electron microscopy [5-10], to ultra-relativistic energies, as high as several GeV in X-ray free electron laser facilities [11,12].

Broader applications of free electron sources include electron beam lithography [13,14], atom-by-atom matter assembly [15], nanoscale radiation sources [16-36] and novel methods of electron microscopy [37-47].

The wide range of free-electron-based applications highlights the importance of developing effective electron wave-shaping techniques, which would enable an even larger design-space in tailoring free-electron-based processes. Free electrons are readily manipulated through electron–light and electron–matter interactions, as manifested by phenomena such as the Kapitza–Dirac effect [48-51] and electron double-slit interference [52,53]. In particular, the structuring of an electron's wavefunction via interference has been experimentally demonstrated [53]. Just as optical waveshaping has uncovered a wealth of novel electromagnetic phenomena [54-59], so electron waveshaping promises to be rich in new electron beam physics and applications. A host of methods has arisen for the design of electron wavepackets, leveraging a variety of mechanisms – including static fields [60-65], radio-frequency cavities [66-71], laser pulses [72-83], and material structures [84,85] – to shape the spatiotemporal profile of an electron pulse, achieving temporal shaping down to the attosecond timescale. Breakthroughs in manipulating the phase structure of electron wavepackets [86,87] have led to further control over properties such as orbital angular momentum [88-90], spin angular momentum [91,92] and propagation trajectory [93,94]. These structured electron beams can be generated through a variety of means including amplitude and phase holograms [95-99], nanoscale magnetic needles [100], and electron–photon interactions [101].

These advances in electron wave-shaping techniques raise the fundamental question of whether quantum electrodynamical (QED) interactions (e.g., light emission) can be controlled via electron wave-shaping. To appreciate the importance of this question in practical applications, consider Bremsstrahlung, the spontaneous emission of a free electron scattering off a static potential. Bremsstrahlung is responsible for the spectrally and angularly broad X-ray background from modern X-ray tubes. If QED interactions can indeed be controlled via electron wave-shaping, Bremsstrahlung could conceivably be made more directional, monochromatic and versatile by structuring the emitting electron wavepacket, analogous to how radio waves are made more directional through structured emitters like phased-array antennas. This effect would be especially exciting in the hard X-ray regime, since the spatial resolution needed to manipulate the phases of hard X-rays cannot be achieved through material fabrication in optical elements, but is readily achievable through electron interference patterns.

In this paper, we present the concept of exploiting quantum interference in QED processes through shaped electron wavepackets, providing a new degree of freedom in the design of these interactions. As an example, we apply our concept to Bremsstrahlung. We show that it is possible to control spontaneous emission from a free electron through quantum interference enabled by electron wave-shaping, just as spontaneous emission from an atom can be controlled through quantum interference between multiple atomic transitions [102] or through multiple atoms, as in superradiance [103]. Specifically, we show that free electron wave-shaping can be used to tailor both the spatial and the spectral distribution of the radiated photons. This results in enhanced directionality, monochromaticity, and versatility of photon emission compared to Bremsstrahlung from an unshaped electron wavefunction. The concept we present can be readily extended to processes involving more massive and non-elementary particles, such as neutrons, whose wavefunction can potentially be shaped as well [104-106].

In a general QED process, the transition probability from input state $|i_1\rangle$ to output state $|f\rangle$ is proportional to $|\langle f|\mathbf{S}|i_1\rangle|^2 \equiv |\delta_1 M_1|^2$, where the S-operator $\mathbf{S}$ transforms the quantum states at the start of the interaction into the quantum states at the end of the interaction. $\delta_1$ is the energy-conserving and/or momentum-conserving Dirac delta distribution, for scenarios with temporal and/or spatial translational invariance accordingly. $M_1$ is the scattering amplitude that abstracts away the part that contains no delta distributions.

We begin by presenting a schematic approach that emphasizes the key points (full details are provided in the concrete examples we consider in Figs. 2 and 3). For an input state $|i_1 + i_2\rangle$, the cross section of the interaction is

$$\text{Cross Section} \propto \int |\delta_1 M_1 + \delta_2 M_2|^2, \tag{1}$$

where the integral in Eq. (1) is carried out over the output states, and $|i_1\rangle \neq |i_2\rangle$. For a general choice of input electron states, the Dirac delta distributions $\delta_1$ and $\delta_2$ peak at different combinations of output particle momenta, resulting in the cross terms cancelling under the integration. i.e., the overall cross section can be written as an incoherent summation of cross sections (Fig. 1a):

$$\text{Incoherent cross section} \propto |M_1|^2 + |M_2|^2, \tag{2}$$

where it is implicit that the various arguments in $M_1$ and $M_2$ have been assigned the values enforced through the integration of the respective Dirac delta distributions.

However, with precision particle wave-shaping, it is possible to select distinct input states such that $\delta_1$ and $\delta_2$ peak at the same combinations of output particle momenta, resulting in quantum interference between the scattering amplitudes associated with $|i_1\rangle$ and $|i_2\rangle$. This interference gives rise to coherent summation (Fig. 1b):

$$\text{Coherent cross section} \propto |M_1 + M_2|^2. \tag{3}$$

which include a nonzero contribution from the cross terms.

We see that the overall cross section is determined not only by the magnitudes of scattering amplitudes $M_1$ and $M_2$, but also by their relative phase, which can be controlled by the relative phase of states $|i_1\rangle$ and $|i_2\rangle$. The scenarios discussed in Eq. (1-3) are readily extended to more than two input states, and reveal the ability of particle wave-shaping to introduce a new degree of freedom in the control of QED processes: namely, the design of QED processes not only through the magnitudes of the constituent scattering amplitudes, but also through the relative phases between these amplitudes.

Our concept of harnessing quantum interference via electron wave-shaping allows us to utilize the relative phases of scattering amplitudes for tailoring QED processes.

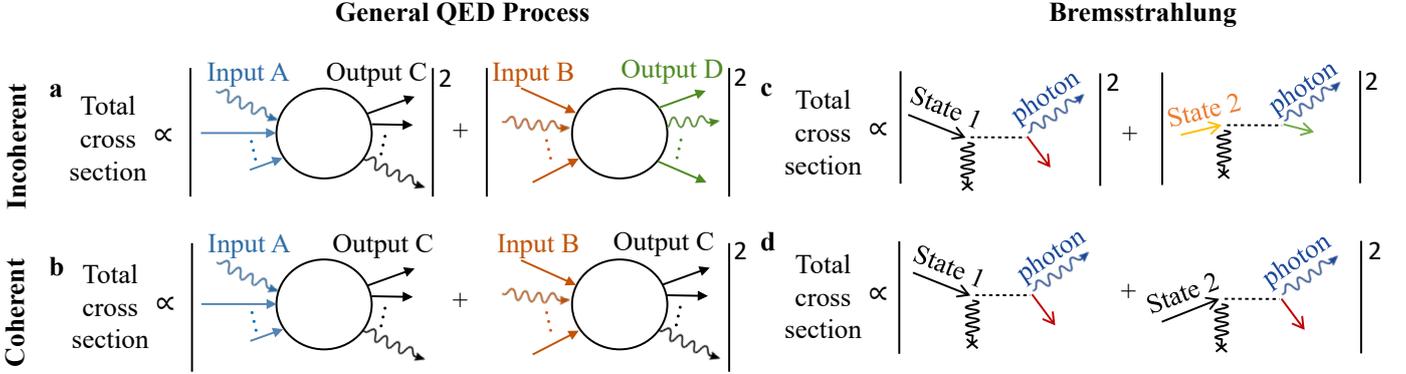

**Figure 1. Coherent and incoherent contributions in quantum electrodynamical (QED) processes.** When the input to a QED process is a superposition of multiple states, e.g., 2 states (Input A and Input B), the overall cross section is typically given by the sum of the cross sections associated with each input state, as in (a). Then, the total cross section is proportional to the sum of the squared-modulus of all the respective scattering amplitudes. However, when the input states are chosen to yield the same output state C, the individual processes coherently interfere, as in (b). The result is a square of summed amplitudes in (b), as opposed to the *sum of squared* amplitudes in (a). Essentially, (a) and (b) illustrate the concept that multiple quantum pathways will add coherently if and only if their output is the same, regardless of how much their input differ from one another. The specific case of Bremsstrahlung is presented in (c) and (d) using Feynman diagrams, illustrating the ideas of (a) and (b) respectively. The diagrams show the spontaneous emission of a photon from a free electron scattering off a static potential. The static potential is represented by the wiggly black line terminating in a cross. The coherent addition in (b) can be harnessed via free electron wave-shaping as a new degree of freedom to tailor the properties of QED processes. In Bremsstrahlung, coherent interference (d) can lead to enhanced directionality, monochromaticity and versatility in the photon output, as explored in Figs. 2 and 3.

Comparing Eqs. (1-3) reveals the conditions to achieve spectrum control via wavefunction interference. This comparison also emphasizes why such possibilities have not been seen before. For example, in [107] the emission of each photon was entangled to an outgoing electron, and the contributions to the emission from different initial electron angles could not interfere because each photon state was entangled to a different outgoing electron state. Similarly, work that considered shaping Cherenkov radiation through the orbital angular momentum (OAM) of electrons [90] found no change to the power spectrum, unless the outgoing electron was post-selected. We attribute these spontaneous emission results to the electron behaving ultimately as a point-like particle (as nicely put by Feynman [108]), regardless of its wavefunction. However, by finding scenarios in which different contributions to an emitted photon state are entangled to the same outgoing electron state, we can achieve interference and a strong dependence on the emitter's wavefunction.

To exemplify the general concept in Figs. 1(a,b), we apply it to Bremsstrahlung, the spontaneous emission of a photon by a free electron scattering off a static potential (Figs. 1(c,d)). We consider two examples for this potential: a neutral Carbon atom in Fig. 2 and the magnetic field of a ferromagnet with nanoscale periodicity (a "nano-undulator") in Fig. 3. The latter case is sometimes referred to as magnetic Bremsstrahlung, or undulator radiation.

Now we consider an input electron wavepacket described as a superposition state $\int \frac{V}{\hbar^3} d^3\boldsymbol{p}\, c_p |p\rangle$ composed of multiple states $|p\rangle$ (labelled by their four-momenta $p$) weighted by complex coefficients $c_p$. We obtain the differential cross section (cross section $\sigma$ per unit angular frequency $\omega_{k'}$ per unit solid angle $\Omega_{k'}$) for an output photon of four-wavevector $k'$ as

$$\frac{d\sigma}{d\omega_{k'} d\Omega_{k'}} = \frac{c}{(2\pi)^4 vT} \int \frac{1}{\hbar^3} d^3\boldsymbol{p}' \left| \int \frac{V}{\hbar^3} d^3\boldsymbol{p} \left[ \delta(E_{p'} + \hbar\omega_{k'} - E_p) \sqrt{\frac{\hbar\omega_{k'}}{8\epsilon_0 E_{p'} E_p}} c_p M_{k'p'p} \right] \right|^2, \quad (4)$$

with

$$M_{k'p'p} = -e^2 \bar{u}(p') \tilde{A}_\nu \left( \frac{\boldsymbol{p}'}{\hbar} + \boldsymbol{k}' - \frac{\boldsymbol{p}}{\hbar} \right) \{ \gamma^\mu \epsilon^*_{k',\mu} i\tilde{S}_F(p' + \hbar k') \gamma^\nu + \gamma^\nu i\tilde{S}_F(p - \hbar k') \gamma^\mu \epsilon^*_{k',\mu} \} u(p). \quad (5)$$

$V$ is the interaction volume, $T$ the interaction time, $e$ the elementary charge, $c$ the speed of light in free space, $v$ the speed of the input particles, $\epsilon_0$ the permittivity of free space, $p^\mu$ ($\mu = 0,1,2,3$) or simply $p$ the four-momentum for electrons (electron energy $E_p \equiv cp^0$), $k^\mu$ or simply $k$ the four-wavevector for photons (angular frequency $\omega_k \equiv ck^0$), and $\epsilon^\mu$ the photon polarization. Bold variables refer to the three-vector counterparts of the respective four-vectors, $\hbar$ is the reduced Planck's constant, $\gamma^\mu$ are the gamma matrices and we use the repeated index convention $k^\mu p_\mu \equiv k^0 p^0 - \boldsymbol{k} \cdot \boldsymbol{p}$. Primes denote variables associated with outgoing particles. Based on the particles' dispersion relations, we have that $p^\mu p_\mu = m^2 c^2$, $k^\mu k_\mu = 0$ and $\epsilon^\mu k_\mu = 0$ ($m$ the electron mass). The $u$-type spinor is given by $u(p) = \begin{bmatrix}\sqrt{p^\mu \sigma_\mu}\,\xi & \sqrt{p^\mu \bar\sigma_\mu}\,\xi\end{bmatrix}^T$, with column vectors $\xi = [1\ 0]^T$ and $\xi = [0\ 1]^T$ corresponding respectively to spin-up and spin-down. $\sigma^\mu = \{1, \sigma_x, \sigma_y, \sigma_z\}$ and $\bar\sigma^\mu = \{1, -\sigma_x, -\sigma_y, -\sigma_z\}$, $\sigma_{x,y,z}$ are 2×2 Pauli matrices. Additionally, $\bar u = u^\dagger \gamma^0$ and $\tilde S_F(p) = (\gamma^\mu p_\mu - mcI)^{-1}$, with $I$ the 4×4 identity matrix. There are two kinds of photon polarizations, given by $\epsilon_1^\mu = \{0,1,0,0\}$ and $\epsilon_2^\mu = \{0,0,1,0\}$ in the case of a photon propagating in the $+z$ direction. In our calculations, we use the metric tensor and gamma matrix conventions of Peskin & Schroeder [109]. $\tilde A^\mu(\boldsymbol{k})$ refers to the Fourier transform of the static potential in the Bremsstrahlung interaction. Besides representing atomic potentials, $\tilde A^\mu(\boldsymbol{k})$ can be used to capture any type of static electromagnetic field, as well as time dependent external fields by making it a function of the full four-vector $k^\mu$.

From the energy-conserving delta distribution in Eq. (4), we see that quantum interference between the processes associated with different input states $p$ would occur if and only if the output of the various processes is identical. This further implies that the input states must have the same energy $E_p$. We obtain the differential cross section in the case of coherent quantum interference as

$$\frac{d\sigma}{d\omega_{k'} d\Omega_{k'}} = \int d\Omega_{p'} \left[\frac{\omega_{k'} |\boldsymbol{p'}|}{8\epsilon_0 (\hbar c)^3 (2\pi)^5 |\boldsymbol{p}|} \left|\int \frac{V}{\hbar^3} d^3\boldsymbol{p}\ c_p M_{k'p'kp}\right|^2\right], \qquad (6)$$

where energy conservation $E_{p'} + \hbar\omega_{k'} - E_p = 0$ is implicitly enforced. Note that the differential rate $d\Gamma/d\omega_{k'} d\Omega_{k'}$ can be obtained from the differential cross section via the relation $d\Gamma/d\omega_{k'} d\Omega_{k'} = (v/V)(d\sigma/d\omega_{k'} d\Omega_{k'})$. We show quantitative results obtained using this formalism in Fig. 2.

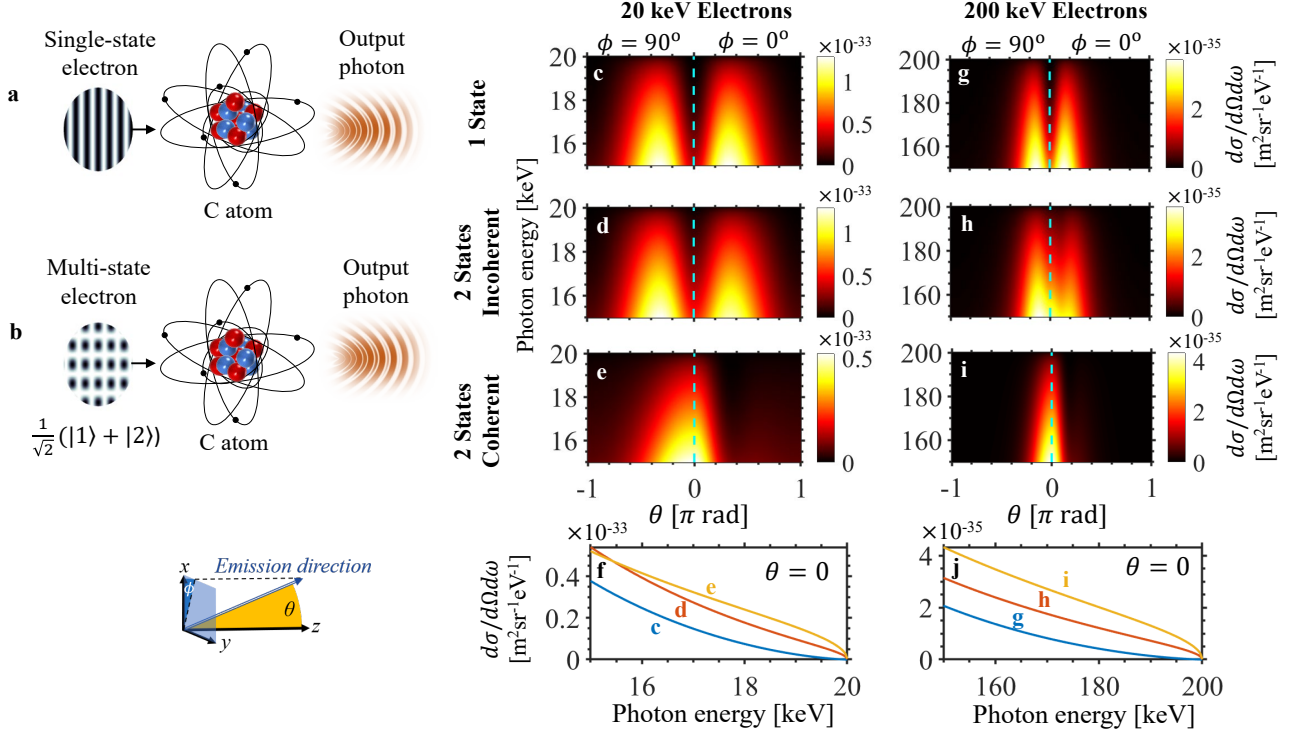

**Figure 2. Enhanced directionality in atomic Bremsstrahlung through shaped electron wavepackets.** In the typical atomic Bremsstrahlung scenario (a), a single momentum state electron scatters off a Carbon atom and emits radiation. Shaping the input electron wavepacket through the use of multiple states as in (b) can enhance the output photon properties through coherent interference between the processes associated with each individual electron state. To illustrate this, (c-e) show the photon output for 20 keV electrons, with a single *z*-directed input electron state in (c), and two input electron states of opposite phase and oriented at $\theta = \pm 15°$ with respect to the *z*-axis in (d) and (e). A donut shaped emission pattern, as indicated by the off-axis peaks, is expected for the single-state (c) and incoherent double-state (d) cases. In contrast, quantum interference between the constituent processes in (e) strongly suppresses off-axis emission, resulting in an emission pattern that is more directional and peaked on-axis. Cross section emission patterns at $\theta = 0$ are compared in (f). The enhanced directionality also holds at other choices of electron energies and angles, as shown in (g-j), which presents the emission spectra corresponding to the scenarios in (c-f) respectively, but for 200 keV electrons, at $\theta = \pm 15°$.

Figure 2 considers Bremsstrahlung radiation where the scattering potential $A^\mu$ is that of a neutral Carbon atom, modeled using a sum of three Yukawa potentials fitted to the results of relativistic Hartree–Fock calculations, which agree well with experimental measurements [110,111] (see Methods). In all cases, the result is averaged over output spin and photon polarization, while the input electron states are taken to be spin-up. Figures 2(a) and (b) illustrates the two scenarios under consideration: an (unshaped) electron state of a single momentum traveling in the +*z* direction, and a shaped electron input obtained by a superposition of 2 states, respectively. In the latter case, each of the two states have probability 0.5, a $\pi$ phase shift with respect to each other, and propagate at ±15º with respect to the +z direction (i.e., shaped input $|i\rangle = (|p_+\rangle - |p_-\rangle)/\sqrt{2}$, where $\boldsymbol{p}_\pm \equiv p_0(\pm\sin\theta_i, 0, \cos\theta_i)$ and $\theta_i = 15º$). Such an input can be realistically generated using holography methods in electron microscopy, with a bi-prism or other analogues of double-slit experiments [53]. Note that the integral over the momentum constituents $\boldsymbol{p}$ of the incoming electron in Eqs. (4) and (6) is treated as a discrete sum over two states in this case. The electron kinetic energy of 20 keV is readily obtained from table-top scanning electron microscopes and from DC electron guns.

Figure 2(c) shows that the emission pattern for the single-state scenario is peaked off-axis (the plot range is cut at the highest possible output photon energy, equals the maximum kinetic energy of the input electron). The comparison of Figs. 2(d-e) show that there is a difference between the two possible emission patterns based on whether or not quantum coherent effects are considered. We see in Fig. 2(e) that the case of coherent superposition leads to a more directional photon output compared to the single-state case in Fig. 2(c) and also compared to the incoherent double-state case in Fig. 2(d). Note that the latter is calculated by summing the cross section results for the two momenta, as would be the case if this superposition state went through decoherence, reducing it to a mixed state of the two momenta of equal probability (i.e. the cross-terms resulting from the squared-modulus in Eq. (6) are ignored). To quantify the increased directionality in the quantum coherent case, we note that the ratio of the on-axis emission to the total emission in the shaped coherent case (Fig. 2(e)) is 3.27, a 12-fold enhancement of the corresponding ratio, 0.273, in the unshaped case (Fig. 2(c)).

Figures 2(g-i) show that the enhanced directionality can also be observed with 200 keV electrons, which are readily obtainable from table-top transmission electron microscopes and radiofrequency electron guns. In this case, the ratio of on-axis emission to total emission in the shaped coherent case (Fig. 2(i)) is 13.04, a 9-fold enhancement of the corresponding ratio, 1.404, in the unshaped case

(Fig. 2(g)). We conclude that shaping the electron wavefunction enables significant control over the output angular distribution, as a direct result of quantum interference between different components of the wavefunction.

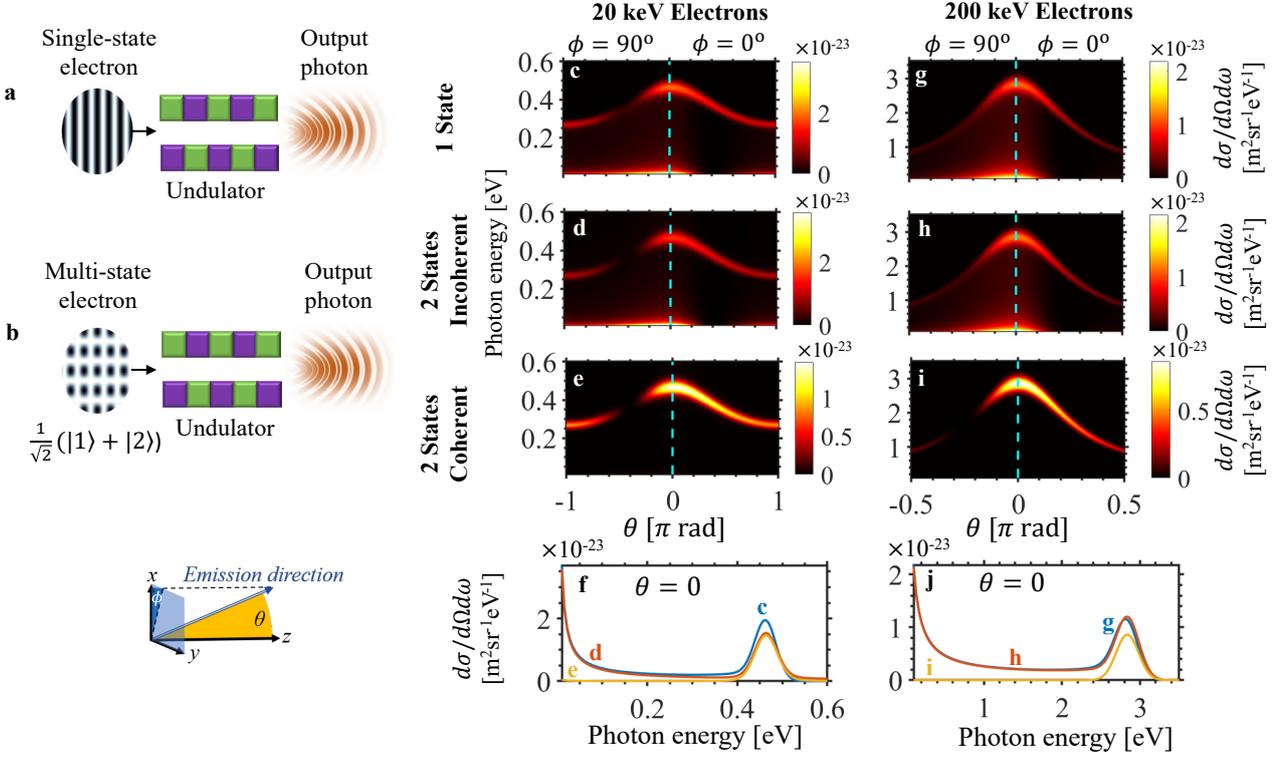

**Figure 3. Enhanced monochromaticity of magnetic Bremsstrahlung (undulator radiation) through shaped electron wavepackets.** We consider the scattering of an input electron off the magnetic field of a nano-undulator, for the case of a single momentum state input electron (a) and that of an input electron made up of two states (b). (c) shows the emission pattern for the single-state input electron scenario in (a). In addition to a relatively monochromatic peak, there is a strong synchrotron radiation-like signature leading to relatively broadband radiation, with significant radiation components at lower photon energies. (d) and (e) show the emission patterns for the double-state input electron scenario in (b), with incoherent and coherent processes considered in (d) and (e) respectively. As (e) shows, the quantum coherence leads to a destructive interference that strongly suppresses the broad synchrotron peak at low photon energies, leading to a more monochromatic output in each direction. Cross sections of the emission patterns at $\theta = 0$ are compared in (f). The suppression of low photon energies continues to hold at other choices of electron energies, as shown in (g-j), which presents the emission spectra for electrons of 200 keV, corresponding to the scenarios in (c-f). The undulator considered is of period 1 μm and has an effective length of 5.3 μm. The two input electron states are of opposite phase and oriented at $\theta = \pm 0.5°$ with respect to the z-axis in (d) and (e), and at $\theta = \pm 0.025°$ in (h) and (i).

Beyond tailoring the *spatial (angular) distribution* of output radiation, quantum interference through electron wave-shaping can also be harnessed to control the *spectral (frequency) distribution* of output photons. This is shown in Fig. 3, which explores Bremsstrahlung from electrons scattering off the fields of a nano-undulator. Few-cycle undulators (also called wigglers) have potential applications in the generation of extremely short pulses of high frequency light [112]. The study of a nano-undulator design is also motivated by recent advances in nanofabrication of magnetic materials that can support large magnetic fields (~1 T) at the surface of nanopatterned ferromagnets [113-115].

The single-state and multi-state electron input scenarios are schematically illustrated in Figs. 3(a,b). The output emission patterns for an electron with a kinetic energy of 20 keV are shown in Figs. 3(c-f). Figure 3(c) corresponds to the single-state input scenario and shows that the main radiation peak in each direction is accompanied by a strong synchrotron radiation-like signature, leading to relatively broadband radiation with significant radiation components at lower photon energies. Figures 3(d-e) correspond to radiation from two crossed electron states propagating at angles ±0.5° with respect to the +z direction, each having probability 0.5. In Fig. 3(d) the emission from the different momenta are summed incoherently, effectively describing a decohered superposition state.

If the two momentum states are coherent, however, interference in the emission channels must be taken into account, as we do in Fig. 3(e). We observe that the radiation profile is significantly modified. In particular, the radiation is much more monochromatic as the synchrotron-radiation-like tail in the low-photon-energy regime is greatly suppressed by destructive interference, whereas the undulator radiation peak remains relatively unaffected. Considering only photon energies above 0.01 eV, the total cross section in the shaped coherent case (Fig. 3e) is 16.5 $\mu m^2$, compared to the total cross section in the unshaped case (Fig. 3(c)) of 24.4 $\mu m^2$, indicating that electron waveshaping has reduced the total photon emission by 67.7%, with the majority of the suppression taking place at off-peak frequencies. To quantify this off-peak suppression, the coherent shaped electron's rate of emission is reduced 71-fold relative to the unshaped electron at a photon energy of 0.01 eV. The enhanced monochromaticity can be directly seen from the photon energy spread (standard deviation) decreasing by more than 10-fold from 62.5% in the unshaped case to 4.3% in the shaped case. In addition to the radiation being more monochromatic, the reduced photon emission at other unwanted frequencies reduces the rate of unwanted energy loss to radiation. Such a dependence on the electron wavepacket has intriguing consequences, as it can potentially lead to a longer mean free path for the electron in matter.

The enhanced monochromaticity induced by quantum interference can also be observed at other electron energies. For instance, Figs. 3(g-j) show the emission patterns for 200 keV input electron momentum states (propagating at angles $\pm 0.025°$ with respect to the $+z$ direction), where we see that the suppression of the broadband synchrotron-radiation-like tail is even more pronounced. Considering only photon energies above 0.1 eV, the total cross section in the shaped coherent case (Fig. 3(i)) is 398.7 $\mu m^2$. We compare this value to the total cross section in the unshaped case (Fig. 3(g)) of 748.4 $\mu m^2$, indicating that electron waveshaping has reduced the total photon emission by 53.3%, with the majority of the suppression taking place at off-peak frequencies. Considering a photon energy of 0.1 eV, we observe a very large off-peak suppression with the shaped coherent photon's rate of emission (Fig. 3(i)) reduced 27,000-fold relative to the case with the unshaped electron (Fig. 3(g)). The enhanced monochromaticity can be directly seen from the photon energy spread (standard deviation) decreasing by more than 10-fold from 65.9% in the unshaped case to 3.6% in the shaped coherent case.

The predictions in our Bremsstrahlung studies (Figs. 2 and 3) can be tested using microscopes with X-ray detectors, e.g. via energy dispersive X-ray spectroscopy (EDS) or electron energy loss spectroscopy (EELS), with the two interfering states created from a single electron state using a bi-prism.

We note that the quantum interferences in QED that arise from the electron wavepacket shaping can be understood as the free-electron QED analogs of coherent interference phenomena in atomic physics. Examples of such coherent phenomena are electromagnetically induced transparency [116], lasing without inversion [117], and refractive index enhancement [118], whereby new physics arises from interference of the transition probability amplitudes between atomic states. Furthermore, the enhancement of radiation using pre-shaped electron wavepackets explored here is highly complementary to other enhancement techniques, including the use of external structures such as photonic crystals [119] and self-induced enhancements like self-amplified spontaneous emission (SASE) [120-124]. Unlike SASE, which involves the bunching of multiple charged particles, our presented mechanism leverages the wave nature of the electron wavepacket instead of the classical distribution. As such, the radiation enhancement we predict can already occur at the level of a single charged particle, and does not require multiple particles. It is noteworthy that just one electron constructed as a superposition of two momentum states can already lead to over 10 times more monochromatic radiation as well as a substantial reduction in unwanted radiation loss.

Our findings provide a definitive answer to the fundamental question: can the quantum nature of the electron wavefunction affect the radiation it emits? When Schrödinger first introduced the quantum wavefunction, he interpreted it as the smooth charge density of a smeared-out particle [125]. Contradictions arising from this view eventually led to the interpretation of the wavefunction as a probability density of a point particle [126]; in the words of Feynman, "The electron is either here, or there, or somewhere else, but wherever it is, it is a point charge" [108]. Yet, intriguingly enough, it had been observed that an electron behaves exactly like a smooth charge density in stimulated emission processes, which have been shown to depend on the waveshape of the emitting electron in both experiment [38] and semiclassical theory [127-129]. However, semiclassical theory does not capture *spontaneous* emission processes (which relies on the quantized nature of light), **and so the fundamental question as to whether electron waveshaping can affect spontaneous emission had remained unanswered.** The significance of a definitive answer to this question has been underscored by recent discussions in the context of shaping electrons for Cherenkov radiation (an example of a spontaneous emission process) [90], and a recent experiment that showed no dependence on wavefunction for Smith-Purcell radiation (yet another spontaneous emission process) in its regime of exploration [107]. Interestingly, recent findings also point out that when the electron is post-selected, spontaneous emission into near-field modes can depend on the symmetry of the initial electron wavefunction [130]. Through a fully quantum theory, we have now shown that fundamental principles support the notion that electron waveshaping *can* affect the emitted radiation. Furthermore, our theory shows that quantum interference (the coherent addition of multiple quantum pathways) is possible under special conditions, and can lead to drastic modifications in the radiation output.

In conclusion, we have presented the concept of engineering quantum interferences in QED processes through shaped electron wavepackets, providing a new degree of freedom in the design and optimization of these processes. As an example, we applied our concept to Bremsstrahlung, showing that it is possible to control this process of spontaneous emission from a free electron through the quantum interference resulting from electron wave-shaping. Specifically, we show that free-electron wave-shaping can be used to tailor both the spatial and the spectral distribution of the radiated photons, enhancing the directionality, monochromaticity, and versatility of photon emission compared to conventional Bremsstrahlung. The reduced photon emission at unwanted frequencies and directions may help to reduce the rate of unwanted energy loss from the radiating electron, and thus potentially lead to a longer mean free path in matter for properly-shaped electrons.

Looking forward, the concept presented in this work can be readily extended to QED processes with other charged particles like protons and ions, as well as processes in other field theories involving more elementary particles, such as pions, muons, and kaons, for which the same principles of wave shaping should apply. The prospect of coherent control over QED processes through particle wave-shaping potentially opens up a wide vista of novel phenomena in fundamental and applied research, where the structure of electron wavepackets provides additional degrees of freedom to control and optimize electron-based quantum processes.

## Methods

**Scalar potential of a neutral atom.** The Carbon atom is modeled using a sum of three Yukawa potentials fitted to the results of the Dirac–Hartree–Fock–Slater (DHFS) self-consistent calculations as described in [110,111]. In the space domain, this potential is given by

$$\phi_{\text{atom}}(\boldsymbol{r}) = \frac{-Ze}{4\pi\epsilon_0|\boldsymbol{r}|} \sum_{j=1}^{2} C_j e^{-\mu_j \frac{|\boldsymbol{r}|}{a_0}}, \quad (M1)$$

where $Z$ is the atomic number, $\epsilon_0$ the permittivity of free space, $a_0$ the Bohr radius and the constants $C_j$ and $\mu_j$ can be obtained from the tables in [110]. For the neutral carbon atom, $Z = 6$, $C_1 = 0.1537$, $C_2 = 0.8463$, $\mu_1 = 8.0404$, and $\mu_2 = 1.4913$. The Fourier transform of Eq. (M1) yields

$$\tilde{\phi}_{\text{atom}}(\boldsymbol{k}) = \frac{-Ze}{\epsilon_0} \sum_{j=1}^{2} \frac{C_j}{|\boldsymbol{k}|^2 + \left(\frac{\mu_j}{a_0}\right)^2}. \quad (M2)$$

The four-vector $\tilde{A}_\nu$ to be used in Eq. (5) is then $\tilde{A}_\nu(\boldsymbol{k}) = \{\tilde{\phi}_{\text{atom}}(\boldsymbol{k}), 0, 0, 0\}$. The scalar potential of this carbon atom in real space is shown in Fig. M1 for the above specified values.

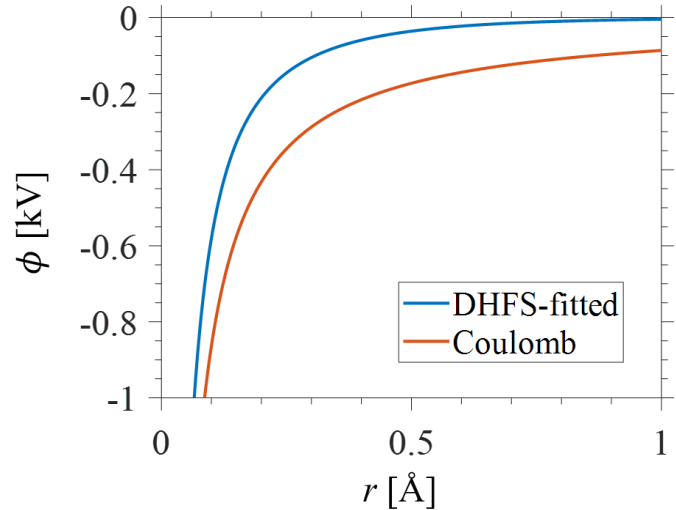

**Figure M1. Electric potential of Carbon atom.** We plot the DHFS-fitted scalar potential (blue) for the C atom used in our atomic Bremsstrahlung calculations, with the Coulomb potential of the unshielded nucleus (red; Eq. M1 with the specified parameters, but setting $C_1 = 1$, $C_2 = \mu_1 = \mu_2 = 0$) for comparison.

**Vector potential of a magnetic undulator.** The expression

$$\mathbf{\Pi}_{\text{und}}(r) = \hat{x}\frac{\Pi_0}{(2\pi)^2}\int\int dk_z dk_y\, f(k_y, k_z) e^{ik_z z} e^{ik_y y} \frac{1}{2}(e^{q_x x} + e^{-q_x x}), \qquad (M3)$$

where $q_x \equiv (k_y^2 + k_z^2)^{\frac{1}{2}}$ and $\Pi_0$ is a constant prefactor, exactly solves the wave equation $\nabla^2 \mathbf{\Pi}_{\text{und}} = 0$, where $\nabla^2$ is the vector Laplacian. Treating $\mathbf{\Pi}_{\text{und}}$ as a Hertz potential (see e.g., [131]), we find that vector potential $\mathbf{A}_{\text{und}} = \mu_0 \nabla \times \mathbf{\Pi}_{\text{und}}$ and the magnetic flux density $\mathbf{B}_{\text{und}} = \mu_0 \nabla \times \nabla \times \mathbf{\Pi}_{\text{und}}$, $\mu_0$ being the permeability of free space. We can verify that $\nabla \times \mathbf{B}_{\text{und}} = \nabla \cdot \mathbf{B}_{\text{und}} = 0$, showing that Eq. (M3) is a valid model for a general static magnetic field in free space. For instance, for the arbitrary scalar function $f = \delta(k_z - k_{z0})\delta(k_y - k_{y0})$, Eq. (M3) gives the potential corresponding to a static magnetic field that is periodic over an infinite area in the $y$ and $z$ dimensions.

The Fourier transform of $\mathbf{A}_{\text{und}}$ is given by:

$$\widetilde{\mathbf{A}}_{\text{und}}(\mathbf{k}) = \mu_0 \Pi_0 f(k_y, k_z)\frac{1}{|\mathbf{k}|^2}[\hat{y}ik_z - \hat{z}ik_y] \times$$

$$\left[e^{\frac{q_x L}{2}}\left(q_x \cos\left(\frac{k_x L}{2}\right) + k_x \sin\left(\frac{k_x L}{2}\right)\right) + e^{-\frac{q_x L}{2}}\left(-q_x \cos\left(\frac{k_x L}{2}\right) + k_x \sin\left(\frac{k_x L}{2}\right)\right)\right], \qquad (M4)$$

where $q_x \equiv (k_z^2 + k_y^2)^{\frac{1}{2}}$ and we terminate the undulator field at $x = \pm L/2$ since otherwise the undulator field blows up at large $x$, which is unphysical. In the $y$ and $z$ dimensions, we have used the profile

$$f(k_y, k_z) = \frac{1}{2}\left[e^{-\frac{(k_z-k_{z0})^2}{\Delta_{k_z}^2} - \frac{(k_y-k_{y0})^2}{\Delta_{k_y}^2}} + e^{-\frac{(k_z+k_{z0})^2}{\Delta_{k_z}^2} - \frac{(k_y+k_{y0})^2}{\Delta_{k_y}^2}}\right]. \qquad (M5)$$

In this study, we chose $k_{y0} = 0$, $k_{z0} = 6.28 \times 10^6$ m$^{-1}$ (corresponding to an undulator period of 1 μm in $z$), $\Delta_{k_y} = \Delta_{k_z} = 6.28 \times 10^5$ m$^{-1}$ (corresponding to 5.3 periods within the full-width-half-maximum of the on-axis magnetic field), and $L = 1$ μm. The magnetic field of this undulator is shown in Fig. M2.

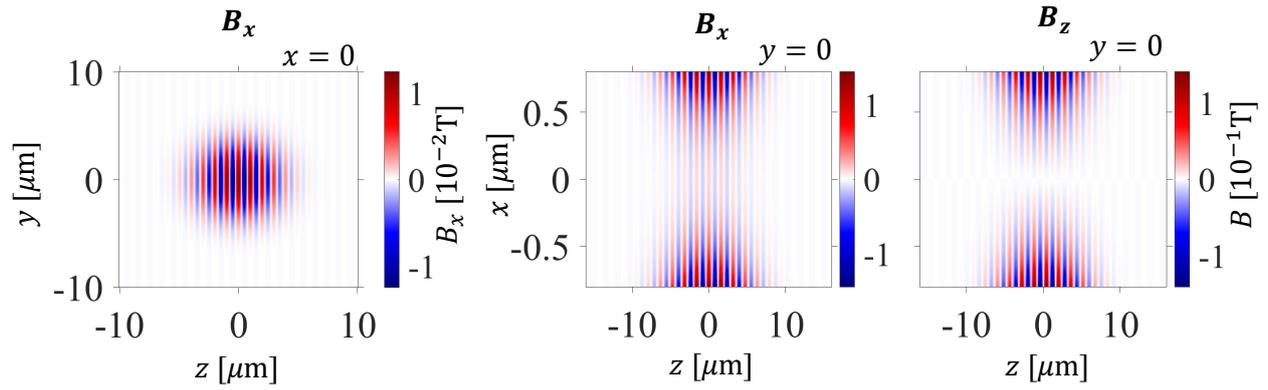

**Figure M2. Magnetic field of nano-undulator.** To visualize the undulator used in our calculations, we plot the corresponding magnetic fields in the $x = 0$ and $y = 0$ planes.

**Data availability**

The data represented in Figs. 2, 3, M1, M2 are available as supplementary information files. All other data that support the plots and other findings within this paper are available from the corresponding authors on reasonable request.

**Code availability**

Not applicable.